\documentstyle[12pt]{article}
\title{{\normalsize
\begin{flushright}SUSX-TH-96-019\\
{\tt hep-ph/9612123}\\
December 1996\\
\end{flushright}}
\vspace{1 cm}
Gauge boson-monopole two particle bound states and duality}
\author{Michael Goodband\thanks{e-mail:
{\tt m.j.goodband@sussex.ac.uk }}
}
\date{}

\begin{document}
\maketitle
\vspace{-24pt}
\begin{center}
{\it
 School of Mathematical and Physical Sciences\\
University of Sussex\\
Brighton BN1 9QH\\
UK
}
\end{center}

\vfill

\begin{abstract}

We consider perturbations in the fields of the 't Hooft-Polyakov monopole, dyon
and point electric solutions. We find a series of bound modes where the fields
are confined to the core of the respective solutions, and these are interpreted 
as bound states of a gauge boson with the respective solutions. We discuss the
spectra of these bound states of the various two particle systems of the 
Yang-Mills-Higgs model in connection with the duality of Sen. 

\end{abstract}
\newpage

\section{Introduction}

In Grand Unified Theories with a simple compact gauge group the series
of symmetry breakings necessary to reach the standard model inevitably
gives rise to monopoles. These monopoles interact with the other 
particles of the theory and the nature of this interaction is of
interest. This interaction has been studied by a number of authors
\cite{Rubakov82}-\cite{Olsen90}. As well as scattering states
\cite{Marciano83}, bound states of bosons and fermions with monopoles
have also been found \cite{Tang82}-\cite{Ajithkumar88}. Perhaps the
most interesting finding is the baryon decay catalysis by a monopole
\cite{Rubakov82}.

Here we are most interested in the work of Sonoda \cite{Sonoda84}. In 
ref.~\cite{Sonoda84} the physics of the massive gauge boson $(W^{\pm})$ 
and the neutral Higgs scalar $(H)$ in the presence of a 't Hooft-Polyakov 
monopole \cite{tHooft74,Polyakov74} was investigated. Spherically symmetric 
perturbations of the fields of the 't Hooft-Polyakov monopole were considered 
and a series of bound states were found. The eigenvalue of the lowest bound 
state was numerically estimated to be $\omega \simeq 0.876 m$ in the 
Prasad-Sommerfield limit ($m$ is the mass of the gauge boson).
We investigate these bound states further by using a more systematic 
approach to the study of general angular momentum perturbations in the fields 
of the 't Hooft-Polyakov monopole. Our motivation for studying them is,
however, quite different.

In ref.~\cite{Montonen77} Montonen and Olive observed that in the Bogomol'nyi
limit \cite{Bogomolnyi76} the masses of all the particles of the theory are
given by the same universal mass formula $m^2 =\eta^2 (q^2 + g^2)$, where
$q$ is the electric and $g$ the magnetic charge of the particle. This formula
applies to both the classical non-singular monopole and dyon solutions as
well as the normal particles, and in part motivated Montonen and Olive to 
conjecture that the $W$-bosons and the monopoles are dual particles. 
The recent construction of supersymmetric models \cite{Seiberg94,Sen94} which 
realize this electromagnetic duality has stimulated a great deal of interest.

We do not consider these supersymmetric models here, however, but consider the
non-supersymmetric model of 't Hooft and Polyakov. We note from the literature
cited above that the massive particles of this theory, whose masses are given 
by the 
universal mass formula, namely the monopole, dyon and $W$-boson form bound 
states with one another. So here we seek to answer the question: are the bound 
state energies of the various two particle systems related by electromagnetic 
duality?

To answer this question we need to find the values of the bound state energies 
in question. The bound state energies for the monopole-dyon system were found 
in ref.~\cite{Gibbons86}. Here we try to complete the picture by using 
perturbative methods to investigate the bound state energies of $W$-bosons in 
the presence of a monopole, a dyon and another $W$-boson.

\section{Yang-Mills-Higgs model}

We are considering field configurations in the Yang-Mills-Higgs theory with
the lagrangian
$$
{\cal L} = -\frac{1}{4} F^a_{\mu\nu} F^{a\mu\nu} + \frac{1}{2}
(D_{\mu}{\rm \Phi}^a)(D^{\mu}\Phi^a)
-\frac{\lambda}{4}(\Phi \cdot \Phi-\eta^2)^2,
$$
where
\begin{eqnarray}
F_{\mu\nu}^a & = & \partial_{\mu}A_{\nu}^a
-\partial_{\nu}A_{\mu}^a
+q\epsilon_{abc}A_{\mu}^bA_{\nu}^c \nonumber \\
D_{\mu} \Phi^a & = & \partial_{\mu}\Phi^a 
+q\epsilon_{abc}A_{\mu}^b \Phi^c, \nonumber
\end{eqnarray}
and $A_{\mu}^a$ and $\Phi^a$ are $SU(2)$ triplets. We will consider
monopole, dyon and point electric solutions to the field equations 
\begin{eqnarray}
D_{\mu}D^{\mu} \Phi^a +\lambda (\Phi \cdot \Phi-\eta^2)
\Phi^a & = & 0 \nonumber \\
D_{\mu}F^{\mu\nu a} + q\epsilon_{abc}\Phi^bD^{\nu}\Phi^c 
& = & 0. \nonumber
\end{eqnarray}
It will be convenient later on if we use dimensionless fields and co-ordinates,
so we make the replacements
$$
r=q\eta R, \hspace{5mm} \phi^a = \eta \Phi^a, \hspace{5mm}
W_{\mu}^a = \eta A_{\mu}^a .
$$
Note that the mass of the massive gauge bosons is $m_w=q\eta $.
With these dimensionless variables we will take the ansatz for the monopole 
with unit magnetic charge to be \cite{tHooft74,Polyakov74}
$$
\phi^a = \hat{r}^a H(r), \hspace{5mm}
W_i^a = -\epsilon_{iab} \frac{\hat{r}^b}{r} W(r),\hspace{5mm}
W_0^a = 0.
$$
The profile functions are solutions to 
\begin{eqnarray}
-\frac{1}{r^2}\frac{d}{dr}\left( r^2\frac{dH}{dr}\right) +2H\frac{(1-W)^2}{r^2}
+\frac{\beta}{2}(H^2 -1)H & = & 0 \nonumber \\
-\frac{d^2 W}{dr^2} + H^2(W-1)+(W-1)\frac{W(W-2)}{r^2} & = & 0, \nonumber
\end{eqnarray}
where we have defined $\beta =2\lambda /q^2$. We will only be considering the
Bogomol'nyi limit $(\beta \rightarrow 0)$ where we have the analytic
Prasad-Sommerfield solution \cite{Prasad75}
$$
H(r) = \frac{\cosh (r)}{\sinh (r)} - \frac{1}{r}, \hspace{5mm}
W(r) = 1 -\frac{r}{\sinh (r)}.
$$
For the dyon with unit magnetic charge we will take the ansatz \cite{Julia75}
$$
\phi^a = \hat{r}^a H(r), \hspace{5mm}
W_i^a = -\epsilon_{iab} \frac{\hat{r}^b}{r} W(r),\hspace{5mm}
W_0^a = \hat{r}^a J(r),
$$
where the profile functions are solutions to 
\begin{eqnarray}
-\frac{1}{r^2}\frac{d}{dr}\left( r^2\frac{dH}{dr} \right) + 
2H\frac{(1-W)^2}{r^2}
+\frac{\beta}{2}(H^2 -1)H & = & 0 \nonumber \\
-\frac{d^2 W}{dr^2} +(H^2-J^2)(W-1)+(W-1)\frac{W(W-2)}{r^2} & = & 0 \nonumber \\
-\frac{1}{r^2}\frac{d}{dr}\left( r^2\frac{dJ}{dr} \right) 
+ 2J\frac{(1-W)^2}{r^2} & = & 0 .\nonumber 
\end{eqnarray}
In the Bogomol'nyi limit we will follow Bogomol'nyi \cite{Bogomolnyi76} and 
take the dyon solution to have the form
$$
\phi^a = \overline{\phi}^a (\sqrt{1-C^2}\: {\bf x} ), \hspace{5mm} 
W_i^a = \sqrt{1-C^2} \: \overline{W}_i^a (\sqrt{1-C^2} \:{\bf x} ),
\hspace{5mm} 
W_0^a = C \: \overline{\phi}^a (\sqrt{1-C^2} \: {\bf x} )
$$
where $(\overline{\phi}^2,\overline{W}_i^a)$ is the Prasad-Sommerfield monopole
solution given above. The parameter $C$ is given by
$$
C=\frac{e}{\sqrt{1+e^2}} \hspace{8mm} \mbox{where} \hspace{8mm}
e=\frac{qQ}{4\pi}
$$
and $Q$ is the electric charge of the dyon. The mass of the dyon in 
dimensionfull units is given by
$$
E= \frac{4\pi \eta}{q} \sqrt{1+e^2},
$$
and setting $e=0$ gives the mass of the monopole. From this we can see that the 
quantity $(1-C^2)^{1/2} \equiv (1-e^2)^{-1/2}$ appearing in the rescaling above 
is the ratio of the monopole and dyon masses. Note that with this form of the 
solution the profile function $J$ is given by $J=CH$.

To investigate the bound state energy of a $W$-boson in the presence of another 
$W$-boson we will consider perturbations about the point electric solution
$$
\phi^3 = (1-\frac{Z}{r} ), \hspace{5mm} W_0^3 = -\frac{Z}{r}.
$$
Substituting this ansatz into the field equations gives
\begin{eqnarray}
D_i^2 \phi^3 & = & 4\pi Z \delta^3 (r) \nonumber \\
D_iG_{0i}^3 & = & 4\pi Z \delta^3 (r) \nonumber
\end{eqnarray}
and so we have a point electric source of strength $4\pi Z$ in rescaled units.
Note that in the rescaled units the $W$-boson has charge 1, and so $Z=1/4\pi$
corresponds to a $W$-boson source.

Now to find the equation for the perturbations 
$(\delta \phi^a,\delta W_{\mu}^a)$ about a solution 
$(\phi^a ,W_{\mu}^a )$ to the field equations, we take 
the second functional derivative of the action. To remove the gauge degrees
of freedom we use the background gauge condition
$$
D^{\mu} \delta W^a_{\mu} + \epsilon^{abc} \phi^b \delta \phi^c = 0.
$$
Imposing this gauge condition, we find the gauge fixed perturbation equations 
are
\begin{eqnarray} \label{pert}
\left( \begin{array}{c}
       (D^2)_{ab} + \delta_{ab} (\frac{\beta}{2} (\phi \cdot \phi -\eta^2 ) 
       +  \phi \cdot \phi ) + (\beta -1) \phi^a \phi^b \hspace{5mm}
        2\epsilon^{abc} D^{\mu} \phi^c \\ 2\epsilon^{abc} D^{\nu} \phi^c 
        \hspace{5mm} 
        g^{\mu\nu} ((D^2)_{ab} + \delta_{ab} \phi \cdot \phi - \phi^a \phi^b ) 
        + 2\epsilon^{abc} G^{\mu\nu c}
       \end{array}  \right) \left( 
       \begin{array}{c}
           \delta \phi^b \\ \delta W_{\mu}^b
        \end{array} \right) & &  \\ 
        & = & 0,\nonumber 
\end{eqnarray}
where the metric is given by $g^{\mu \nu} = diag (-1,1,1,1)$.
The gauge condition does not, however, fix the gauge completely. To remove the
remaining degrees of freedom requires the introduction of Fadeev-Popov ghosts
$\eta^a$ which satisfy
\begin{eqnarray}
\left( (D^2)_{ab} + \delta_{ab} \phi \cdot \phi - \phi^a \phi^b \right) 
\eta^b  =  0 \label{FP}. 
\end{eqnarray}
If we count the number of degrees of freedom we find that equation \ref{pert} 
has 15 and equation \ref{FP} has 3. When calculating any quantum corrections 
the Fadeev-Popov ghost contribution appears with a factor of $-2$ relative to 
the contribution from the modes of equation \ref{pert}. In this way the 
Fadeev-Popov ghosts cancel 6 degrees of freedom of equation \ref{pert} leaving 
9 physical degrees of freedom. They correspond to 6 degrees of freedom for the 
2 massive gauge bosons, 2 degrees of freedom for the massless gauge boson and 
one degree of freedom for the Higgs boson.

We seek to solve equation \ref{pert} for the physical bound states for general
angular momentum in the presence of the three solutions given above. To check
that the states we find are actually physical states we shall also solve 
equation \ref{FP}. If the eigenspectrums of equation \ref{FP} correspond to an 
eigenspectrum of equation \ref{pert} then that eigenspectrum is not physical; 
it is cancelled by the ghosts.

\section{Perturbations about the monopole}

If we substitute the monopole solution into the gauge fixed second variation
of the action, we find
\begin{eqnarray}
\delta S_{gf} & = & \frac{\eta}{q}
\int r^2 dr \int d \Omega \left( \phi^a \left[ D^2_{ab} + 
 \delta^{ab} \left( \frac{\beta}{2}(H^2 -1) + H^2 \right)  
 + \hat{r}^a \hat{r}^b(\beta -1)H^2 \right] \phi^b 
  \right. \nonumber \\
  & - & \left. W_0^a \left[ D^2_{ab}+H^2(\delta^{ab}-\hat{r}^a \hat{r}^b)
  \right] W_0^b + W_i^a \left[ D^2_{ab}+H^2(\delta^{ab}-\hat{r}^a \hat{r}^b)
  \right] W_i^b \right. \nonumber \\
 & + & \left. 4\epsilon^{abc} \phi^a W_i^b (D_i \phi )^c 
 + 2\epsilon^{abc}  W_i^a W_j^b G_{ij}^c \right),
\end{eqnarray}
and the Fadeev-Popov ghost term is
\begin{eqnarray}
\delta S_{FP} & = & \frac{\eta}{q}
\int r^2 dr \int d \Omega  \eta^a \left[ D^2_{ab}
+H^2(\delta^{ab}-\hat{r}^a \hat{r}^b) \right] \eta^b , 
\end{eqnarray}
where 
\begin{eqnarray}
(D^2 )_{ab} & = & \delta_{ab} \partial_0^2 + \delta_{ab} 
\left( -\nabla_r^2 + \frac{L^2}{r^2} 
+ \frac{W^2}{r^2} \right)  - 2\frac{W}{r^2} \epsilon_{iab} L_i 
+ \hat{r}^a \hat{r}^b \frac{W^2}{r^2}  \nonumber \\
G_{ij}^c & = & \epsilon^{kij} \left[ \hat{r}^k \hat{r}^c 
\left( \frac{W^{\prime}}{r} + \frac{W}{r^2} (W-2) \right)  
- \delta^{kc}\frac{W^{\prime}}{r}  \right] \nonumber \\
(D_i \phi )^c & = &  \hat{r}^i \hat{r}^c\left( H^{\prime}-\frac{H}{r}(1-W)
\right) + \delta^{ic} \frac{H}{r} (1-W), \nonumber
\end{eqnarray}
and $L$ is the angular momentum operator $L=-i\hat{r}\times \nabla$.
Note that because the monopole solution is time independent and $W_0^a =0$ we 
can separate out the time derivatives and set up a standard eigenvalue problem.

The monopole solution is invariant under rotations generated by
$K=L+T+S$, and so we seek to expand the perturbations as eigenstates of $K$. 
First we consider the scalar fields $(W_{0}^a,\eta^a,\phi^a )$ which have spin 
zero and so $K=L+T$. Simultaneous eigenfunctions of $K^2$, $K_3$, $L^2$, and 
$T^2$ can be constructed from eigenfunctions of $L^2$, $L_3$, $T^2$ and $T_3$ 
by using standard Clebsch-Gordan methodology
$$
\Phi^{km}_{lt} = \sum_{m_l m_t} {\cal C}^{km}_{ltm_lm_t} Y_{lm_l} Y_{tm_t}
\hspace{5mm} \mbox{where} \hspace{5mm} m=m_l + m_t 
$$
where ${\cal C}$ are Clebsch-Gordan coefficients and $Y_{lm_l}$ are the 
spherical harmonics. We will, however follow ref.~\cite{Carson90} and define 
vector spherical harmonics in terms of operators acting on the spherical 
harmonics by
$$
Y^a_{kjm} = N_{k,j}^{-1/2} O_j^a Y_{k,m}.
$$
The vector operators $O_j$ are given by
\begin{eqnarray}
O_0 & = & iL \nonumber \\
O_1 & = & \hat{r} (\overline{L} +1) + i\hat{r}\times L \nonumber \\
O_{-1} & = & \hat{r}\overline{L} - i\hat{r}\times L \nonumber 
\end{eqnarray}
where
$$
\overline{L} \equiv \left(L^2 +\frac{1}{4} \right)^{1/2} -\frac{1}{2}.
$$
The normalization constants $N_{k,j}$ are given by the orthogonality condition
$$
\int d\Omega (Y^a_{kjm})^{\dagger} Y^a_{k^{\prime}j^{\prime}m^{\prime}} =
\delta_{kk^{\prime}} \delta_{jj^{\prime}} \delta_{mm^{\prime}}.
$$
The vector spherical harmonics $Y^a_{kjm}$ defined above are eigenfunctions of
$K^2$, $K_3$, $L^2$, and $T^2$
\begin{eqnarray}
K^2 Y^{a}_{kjm} & = & k(k+1) Y^{a}_{kjm} \nonumber \\
K_3 Y^{a}_{kjm} & = & m Y^{a}_{kjm} \nonumber \\
L^2 Y^{a}_{kjm} & = & (k+j)(k+j+1) Y^{a}_{kjm} \nonumber \\
T^2 Y^{a}_{kjm} & = & 2 Y^{a}_{kjm} \nonumber
\end{eqnarray}
We now expand the scalar fields as
\begin{eqnarray}
\phi^a & = & \sum_{km} \sum_{j=-1}^{1} \phi^{km}_{j} Y^a_{kjm} \nonumber \\
\eta^a & = & \sum_{km} \sum_{j=-1}^{1} \eta^{km}_{j} Y^a_{kjm} \nonumber \\
W_0^a & = & \sum_{km} \sum_{j=-1}^{1} \tilde{W}^{km}_{j} Y^a_{kjm}. \nonumber \\
\end{eqnarray}
Note that the orbital angular momentum $l$ is given by $l=k+j$, and so for 
each value of $k$ there will be three values of $l$. So the subscript $j$ on
the vector spherical harmonics denotes the three different values of $l$ that
are possible for each value of $k$.

For the spatial components of the gauge field we have non-zero spin and isospin
and so $K=L+T+S$. There are two different ways of adding three angular momenta,
differing in the order in which the addition is performed. We choose to
first construct eigenvectors of
$$
J=T+S.
$$
Note that $J$ here is not the total angular momentum operator. For spin one and 
isospin one there are three possible values of $j$, $j=0,1,2$ denoting scalar, 
pseudovector and tensor representations of $J$.
If we then consider eigenstates of 
$$
K=J+L = T+S+L
$$
we can see that in general for each $k$; for $j=0$ there is only one value of 
$l$, $l=k$; for $j=1$ there are three values of $l$, $l=|k-1|,k,k+1$;
and for $j=2$ there are five values of $l$, $l=|k-2|,|k-1|,k,k+1,k+2$.  
So we introduce the variables $j_1$ and $j_2$ where $j_1$ is the
$J$ eigenvalue and $j_2$ takes the values $j_2 = -j_1, \ldots ,j_1$ so that in 
general, $l$ is given by
$$
l=k+j_2.
$$
With this notation we expand $W_i^a$ in tensor spherical harmonics as
$$
W_i^a  =  \sum_{km} \sum_{j_1=0}^{2} \sum_{j_2=-j_1}^{j_1} 
W^{km}_{j_1j_2} Y^{ai}_{kj_1j_2m} 
$$
where the tensor spherical harmonics $Y^{ai}_{kj_1j_2m}$ are defined in terms of
operators acting on spherical harmonics by 
$$
Y^{ai}_{kj_1j_2m} = N_{kj_1j_2}^{-1/2} O^{ai}_{j_1j_2} Y_{km},
$$
where the tensor operators $O_{j_1j_2}$ are constructed from the $O_j$.
The tensor spherical harmonics defined in this way are
eigenfunctions of $K^2$, $K_3$, $L^2$ and $J^2$
\begin{eqnarray}
K^2 Y^{ai}_{kj_1j_2m} & = & k(k+1) Y^{ai}_{kj_1j_2m} \nonumber \\
K_3 Y^{ai}_{kj_1j_2m} & = & m Y^{ai}_{kj_1j_2m} \nonumber \\
L^2 Y^{ai}_{kj_1j_2m} & = & (k+j_2)(k+j_2+1) Y^{ai}_{kj_1j_2m} \nonumber \\
J^2 Y^{ai}_{kj_1j_2m} & = & (T+S)^2 Y^{ai}_{kj_1j_2m} =
j_1 (j_1 +1) Y^{ai}_{kj_1j_2m}. \nonumber
\end{eqnarray}
Note that the scalar ($j_1=0$), pseudovector ($j_1=1$) and tensor ($j_1=2$)
representations of $J$ have all been incorporated into the single set of 
functions $Y^{ai}_{kj_1j_2m}$.

Now the monopole is also invariant under the action of the parity operator $P$.
If we apply the parity operator to the perturbations, then from
$$
P Y^{a}_{kjm} = (-1)^{k+j} Y^{a}_{kjm} \hspace{5mm} \mbox{and} \hspace{5mm}
P Y^{ai}_{kj_1j_2m} = (-1)^{k+j_2} Y^{ai}_{kj_1j_2m}
$$
we find that the perturbations separate into two channels under parity given by
$$
\psi^+_{km} = \left( \begin{array}{c}
                W^{km}_{22} \\ W^{km}_{20} \\ W^{km}_{2-2} \\ W^{km}_{10} \\
                W^{km}_{00} \\ \phi^{km}_{0}
                \end{array} \right), \hspace{20mm}
\psi^-_{km} = \left( \begin{array}{c}
                W^{km}_{21} \\ W^{km}_{2-1} \\ W^{km}_{11} \\ W^{km}_{1-1} \\
                \phi^{km}_{1} \\ \phi^{km}_{-1}
                \end{array} \right)
$$
where the superscripts are such that
\begin{eqnarray}
+ & & \hspace{5mm} \mbox{denotes parity} \hspace{5mm} (-1)^k \nonumber \\
- & & \hspace{5mm} \mbox{denotes parity} \hspace{5mm} (-1)^{k+1}.\nonumber
\end{eqnarray}
The perturbation equations for $\psi_{km}^{\pm}$ decouple. Note that the vector
and tensor spherical harmonics are eigenstates of $K^2$, $K_3$, $L^2$ and $J^2$
and so are not eigenstates of the charge operator
$$
\hat{Q} = \hat{r}.T.
$$
Substituting these expansions into $\delta S_{gf}$ and performing the 
angular integrals gives the perturbation operator 
\begin{eqnarray}
M_{\alpha \beta}^{(\pm)} & = & \delta_{\alpha \beta} \left[
-\nabla_r^2 + \frac{(k+j_2 )(k+j_2 +1)}{r^2} + \frac{W^2}{r^2} + H^2
+\frac{\beta}{2} (H^2 -1) P_{\phi} \right] \nonumber \\
 & + & 2\frac{W}{r^2} C_{\alpha\beta}^{(\pm )(1)}
 + \left( \frac{W^2}{r^2} - H^2 + \beta H^2 P_{\phi} \right)
 C_{\alpha\beta}^{(\pm )(2)} \nonumber \\
 & + & 2\frac{W^{\prime}}{r} C_{\alpha\beta}^{(\pm )(3)}
  +  2 \frac{W}{r^2} (W-2) C_{\alpha\beta}^{(\pm )(4)} \nonumber \\
 & + & 4\frac{H}{r} (1-W) C_{\alpha\beta}^{(\pm)(5)}
 + 4 \left( H^{\prime} - \frac{H}{r} (1-W) \right) 
 C_{\alpha\beta}^{(\pm )(6)} \nonumber
\end{eqnarray}
where $(\pm )$ denotes the two channels with different parity, the
subscripts $\alpha$ and $\beta$ denote the modes labeled by $j_1$
and $j_2$ (i.e. $\alpha = (j_1,j_2)$ and 
$\beta = (j_1^{\prime},j_2^{\prime})$), and $P_{\phi}$ is a projection operator
which is 1 for a $\phi $ field and 0 for the $W$ fields. The
$C_{\alpha\beta}^{(\pm )(1-6)}$ are symmetric matrices of coefficients 
resulting from the integration over angular co-ordinates. This notation is 
chosen to be the same as that used in ref.~\cite{Carson90}, and the matrix 
elements of $C_{\alpha\beta}^{(\pm )(1-6)}$ are taken from there.

The perturbations for the time components (and the ghosts) similarly separate
under parity
$$
\tilde{W}_{km}^+ = \left( \begin{array}{c}
                       \tilde{W}_{0}^{km} 
                       \end{array} \right) \hspace{5mm}
\tilde{W}_{km}^- = \left( \begin{array}{c}
                       \tilde{W}_{1}^{km} \\ \tilde{W}_{-1}^{km}
                       \end{array} \right) \hspace{20mm}
$$
where the superscripts $\pm$ denote the parities as before.

The perturbation operator for the time components (and the ghosts) is
\begin{eqnarray}
\tilde{M}_{\alpha\beta}^{(\pm)} & = & \delta_{\alpha\beta}
\left[ -\nabla_r^2 + \frac{(k+j_2)(k+j_2+1)}{r^2} + \frac{W^2}{r^2} + H^2
\right] \nonumber \\
& +& \frac{2W}{r^2} C_{\alpha\beta}^{(\pm)(1)} + \left( \frac{W^2}{r^2} - H^2
\right) C_{\alpha\beta}^{(\pm)(2)} \nonumber
\end{eqnarray}
where the $C_{\alpha\beta}^{(\pm)(1,2)}$ elements are the same as those for
the $\phi$ terms.

\subsection*{$k=0$ modes}

If we turn our attention to the $k=0$ modes we find that, because of the 
constraints $l=k+j_2 \geq 0$ and $k \geq 0$, the number of channels for $k=0$ 
are reduced. For $k=0$ the allowed perturbations are
$$
\psi_{00}^+ = \left( \begin{array}{c}
                     W_{22}^{00} \\  W_{00}^{00}
                     \end{array} \right)  \hspace{15mm}
\psi_{00}^- = \left( \begin{array}{c}
                     W_{11}^{00} \\  \phi_{1}^{00}
                     \end{array} \right). 
$$
The $\psi_{00}^+$ mode satisfies the perturbation equation
\begin{equation} \label{+zero}
\left( \begin{array}{cc}
        M_{11} & M_{12} \\ M_{21} & M_{22}
        \end{array}  \right) \left(  \begin{array}{c}
                                      W_{22}^{00} \\ W_{00}^{00}
                                     \end{array} \right) = \omega^2 
        \left(  \begin{array}{c}
                W_{22}^{00} \\ W_{00}^{00}
                  \end{array} \right),
\end{equation}
where
\begin{eqnarray}
M_{11} & = & -\nabla_r^2 + \frac{6}{r^2} - 6\frac{W}{r^2} + \frac{5}{3} 
\frac{W^2}{r^2} + \frac{1}{3} H^2 + \frac{8}{3} \frac{W^{\prime}}{r}
+\frac{2}{3} \frac{W(W-2)}{r^2} \nonumber \\
M_{22} & = & -\nabla_r^2 + \frac{4}{3} \frac{W^2}{r^2} + \frac{2}{3} H^2 
- \frac{8}{3} \frac{W^{\prime}}{r}
+\frac{4}{3} \frac{W(W-2)}{r^2} \nonumber \\
M_{12} & = & M_{21} = \frac{\sqrt{2}}{3} \left( \frac{W^2}{r^2} -H^2 \right)
- 2 \frac{\sqrt{2}}{3} \left( \frac{W^{\prime}}{r} + \frac{W(W-2)}{r^2} 
\right), \nonumber
\end{eqnarray}
and the $\psi_{00}^-$ mode satisfies the perturbation equation
$$
\left(  \begin{array}{c}
        -\nabla_r^2 + 2 \frac{(1-W)^2}{r^2} + H^2 + \frac{W(W-2)}{r^2}
        \hspace{10mm} 2\sqrt{2} \frac{H}{r} (1-W) \\
         2\sqrt{2} \frac{H}{r} (1-W)  \hspace{10mm} 
         -\nabla_r^2 + 2 \frac{(1-W)^2}{r^2} + \frac{\beta}{2} (3H^2 -1)
         \end{array} \right) \left( \begin{array}{c}
                                    W_{11}^{00} \\ \phi_{1}^{00}
                                    \end{array} \right) = \omega^2
                            \left( \begin{array}{c}
                                    W_{11}^{00} \\ \phi_{1}^{00}
                                    \end{array} \right).
$$
If we substitute in a numerical solution for the profiles $H$ and $W$, these
equations can be solved numerically for general $\beta$ by standard matrix 
methods. The most interesting case is the Prasad-Sommerfield limit where we
have an analytic solution. Despite the existence of the analytic solution we 
have been unable to solve the perturbation equations analytically. 
We can however, proceed analytically if we take the far field forms of the
profile functions
$$
H \rightarrow 1 -\frac{1}{r}, \hspace{5mm}
W \rightarrow 1 \hspace{5mm} \mbox{for} \hspace{5mm}
r \rightarrow \infty .
$$
Substituting this into  the perturbation equation for $\psi_{00}^+$ gives
for $r \rightarrow \infty $
$$
\left( \begin{array}{cc}
       -\nabla_r^2 -\frac{2}{3r} + \frac{4}{3r^2} + \frac{1}{3} & 
       \frac{\sqrt{2}}{3} \left( -1 + \frac{2}{r} +\frac{2}{r^2} \right) \\
        \frac{\sqrt{2}}{3} \left( -1 + \frac{2}{r} +\frac{2}{r^2} \right) 
       & -\nabla_r^2 -\frac{4}{3r} + \frac{2}{3r^2} + \frac{2}{3}
       \end{array} \right) \left( \begin{array}{c}
                                  W_{22}^{00} \\ W_{00}^{00}
                                  \end{array} \right) = \omega^2
                           \left( \begin{array}{c}
                                  W_{22}^{00} \\ W_{00}^{00}
                                  \end{array} \right), 
$$    
for which we can find two orthogonal eigenvectors. One of the eigenvectors is 
$$
\left( \begin{array}{c}
      W_{22}^{00} \\ W_{00}^{00}
       \end{array} \right) = \left( \begin{array}{c}
                       1 \\ 1/\sqrt{2}
                    \end{array} \right) w_1
$$
where the function $w_1(r)$ satisfies
$$
\left( -\nabla_r^2 + \frac{2}{r^2} \right) w_1 = \omega^2 w_1.
$$
This has spherical Bessel function solutions for all $\omega^2 \geq 0$. The 
other eigenvector is
$$
\left( \begin{array}{c}
      W_{22}^{00} \\ W_{00}^{00}
       \end{array} \right) = \left( \begin{array}{c}
                       1 \\ -\sqrt{2}
                    \end{array} \right) w_2
$$
and the function $w_2(r)$ satisfies
$$
\left( -\nabla_r^2 + 1 - \frac{2}{r} \right) w_2 = \omega^2 w_2.
$$
If we write $w_2(r) =u(r)/r$, rescale $\rho =2(1-\omega^2)^{1/2} r$ and define
$\Lambda=1/(1-\omega^2)^{1/2}$ then we obtain Whittakers form of the 
hypergeometric equation
$$
\left( \frac{d^2}{d\rho^2} - \frac{1}{4} + \frac{\Lambda}{\rho} + 
\frac{\frac{1}{4}-\mu^2}{\rho^2} \right) u(\rho ) =0,
$$
with $\mu^2 =1/4$. For $\omega^2 \geq 1$ this equation has Whittaker function
solutions $W_{\Lambda,\mu}(\rho)$ for all $\Lambda$ which are oscillatory at 
large $\rho$. For $1 > \omega^2 \geq 0$ the Whittaker function solutions
only exist for $\Lambda = 1,2,3,\ldots$ and the solution decays exponentially 
at large $\rho$. These bound states are analogous to the states of the 
non-relativistic hydrogen atom, and have eigenvalues
$$
\omega_n^2 = 1 - \frac{1}{n^2} \hspace{5mm} \mbox{where} \hspace{5mm}
n=1,2,3,\ldots
$$
The Whittaker function solutions can actually be written as Laguerre 
polynomials, since if we set $u=\rho {\rm e}^{-\rho/2}v$ we get Laguerre's 
equation
\begin{equation}
\left( \rho \frac{d^2}{d\rho^2} + (2-\rho)\frac{d}{d\rho} + (\Lambda -1)
\right) v(\rho) =0,
\end{equation}
which for $\Lambda = n \in {\cal Z}$ has associated Laguerre polynomial
solutions $L_n^1 (\rho)$.

If we now consider the equation for $\psi_{00}^-$ in the limit 
$r \rightarrow \infty$ we find
$$
\left( \begin{array}{cc}
       -\nabla_r^2 + 1 - \frac{2}{r} & 0 \\
       0 & -\nabla_r^2 
       \end{array} \right) \left( \begin{array}{c}
                                  W_{11}^{00} \\ \phi_{1}^{00}
                                  \end{array} \right) = \omega^2
                           \left( \begin{array}{c}
                                  W_{11}^{00} \\ \phi_{1}^{00}
                                  \end{array} \right). 
$$
The Higgs scalar field has spherical Bessel function solutions at large $r$,
while the $W_{11}^{00}$ field has the same type of bound solutions as above
with eigenvalues
$$
\omega_n^2 = 1 - \frac{1}{n^2} \hspace{5mm} \mbox{where} \hspace{5mm}
n=1,2,3,\ldots
$$

These spherical Bessel function and Whittaker function solutions are finite at 
the origin and so are physically acceptable solutions for all $r$. They were
obtained using the far field limit for the profiles which ignores any effect of
the monopole core, so it is not obvious that the eigenvalues obtained will still
hold when we use the full analytic solution.
However, numerical solution of the perturbation equations with the full analytic
form of the profile functions gives numerical results which are consistent with 
the analytic result obtained in the far field limit.

Table 5 shows the analytic and numerical results for the 
eigenvalues of the lowest identifiable bound modes. The numerical results were 
obtained by solving the $\psi_{00}^+$ equation using standard matrix methods. 
The non-linear map $s=\tanh (r)$ was used to assign more lattice points to the 
core of the monopole where the fields are varying most. Apart from the $n=2$ 
eigenvalue we can see that the agreement between the analytic and numerical 
results is fairly good. In trying to argue that the numerical results actually 
agree with the analytic result we have to address two principle shortcomings of 
the numerical results: the difference between the $n=2$ numerical and analytic 
results; and the limited number of bound modes identified.

The $n=2$ bound mode differs from the other bound modes in that the fields of 
this mode vary most in exactly the same place that the monopole profiles vary
most. We would argue that this leads to the error in this mode being greater 
than for the other modes. In fact we found a similar situation in 
ref.~\cite{Goodband95a} where we considered perturbations about a Nielsen-Olesen
vortex \cite{Nielsen73}. There we found a similar bound mode whose fields varied
most in exactly the same place as the string profiles, and its eigenvalue was
the most sensitive of all the modes to the accuracy of the string profiles
used. 

This $n=2$ mode also appears to be the one greatest affected by the omission of
the origin, which is necessary because of the $1/r^2$ centrifugal terms in the
perturbation equation. The effect that the omission of the origin can have is
best illustrated by considering the analogous case of perturbations about the
kink \cite{Rajaraman82}. In this case the perturbation equation can be solved
analytically and two discrete modes are found. One is a $\omega^2 =0$ 
translation mode and the other is a $\omega^2 =0.75\; m^2$ bound mode ($m$ is 
the
mass of the Higgs boson in this case). If we solve this eigenproblem on a 
lattice numerically we obtain agreement with the analytic values for the 
discrete modes above. However, if we remove the lattice point at the origin, 
there is a $0.01\; m^2$ shift in the eigenvalue of the bound mode, whereas the 
$\omega^2=0$ translation mode eigenvalue is unaltered to order $0.01\; m^2$.

We would therefore argue that these affects, notably the omission of the origin,
are responsible for the disagreement between the numerical and analytic results
of the $n=2$ mode, and that its true eigenvalue is in fact given by the
analytic result.

Now the limit on the number of bound modes that could be identified in the 
numerical procedure is not really surprising. In the analytic expression
the difference between the eigenvalues of adjacent modes decreases for the 
higher modes, whereas in solving the eigenvalue problem numerically the 
differences between eigenvalues that can be resolved is dependent on the number 
of lattice points and the box size. Our use of the non-linear $\tanh$ map was
such that the number of lattice points and box size were increased 
simultaneously. We can see that the eigenvalues of the $n=1,\ldots,5$ bound
modes do not change significantly as the number of lattice points is increased.
The eigenvalue of the $n=6$ bound mode, the highest bound mode identifiable with
a 100 point lattice, does change significantly as the number of lattice points
is increased; it tends towards the analytic result. The higher the eigenvalue
of the bound mode, the further from the monopole core the fields of the bound
mode extend, hence as the box size is increased (note that box size and lattice
size increase simultaneously) the number of identifiable bound modes increases.
We would therefore expect that if we continued to increase the number of lattice
points (and box size) we would continue to find more bound modes, and their
eigenvalues would agree with the analytic result. It is worth noting the $1/r$
dependence of the monopole profile $H$ for large $r$, which in a sense means 
that all box sizes are too small.

As for the interpretation of the modes, for $\psi_{00}^+$ we can interpret the 
two eigenspectrums as being associated with the massless gauge boson and one of 
the massive gauge bosons. The massive gauge boson possesses bound states. For 
$\psi_{00}^-$ we can interpret the two eigenspectrums as being associated with 
the massless Higgs boson and the other massive gauge boson. It is again the 
massive gauge boson which has bound states.

To check that these modes are in fact physical we solve the equation for the 
ghosts. For $k=0$ the only allowed mode is
$$
\eta_{00}^- = \eta_{1}^{00}
$$
which satisfies the equation
\begin{equation} \label{Laguerre}
\left[ -\nabla_r^2 + \frac{2(1-W)^2}{r^2} \right] \eta_{1}^{00}
=\omega^2 \eta_{1}^{00}.
\end{equation}
This has no bound mode solutions and so the bound states found above are 
physical. This ghost mode cancels the $\tilde{W}^{00}_{1}$ and $\phi^{00}_{1}$ 
massless modes.

These results agree with ref.~\cite{Sonoda84} where he only studied the 
$\psi_{00}^+$ perturbations for $k=0$ and not the $\psi_{00}^-$ type. He found 
an infinite number of bound states and numerically estimated the lowest to be 
$\omega^2 \simeq 0.768$. This agrees well with the value we obtained numerically
for the first non-zero eigenvalue, but he did not find the bound zero mode.
This spherically symmetric zero mode is not canceled by the ghosts, and is
in fact associated with a physical degree of freedom of the monopole.
Finally no negative modes were found in agreement with the perturbative
stability analysis of ref.~\cite{Yoneya77}

\section{Perturbations about the dyon}

Substituting the dyon solution into the gauge fixed second variation
of the action gives
\begin{eqnarray}
\delta S_{gf} & = & \frac{\eta}{q}
\int r^2 dr \int d \Omega \left( \phi^a \left[ (D^2)_{ab}
+ \delta^{ab} \left(\frac{\beta}{2}(H^2 -1) + H^2 \right) 
+ \hat{r}^a \hat{r}^b (\beta -1)H^2  \right] \phi^b \right. \nonumber \\
 & - & \left. W_0^a \left[ (D^2)_{ab} + H^2 (\delta^{ab} -\hat{r}^a \hat{r}^b)
 \right] W_0^b + W_i^a \left[ (D^2)_{ab} + H^2 
 (\delta^{ab} -\hat{r}^a \hat{r}^b) \right] W_i^b \right. \nonumber \\
 & + & \left. 4\epsilon^{abc} \phi^a W_i^b (D_i \phi )^c 
 + 4\epsilon^{abc} W_0^a W_i^b G_{0i}^c + 2\epsilon^{abc} W_i^a W_j^b G_{ij}^c
 \right),
\end{eqnarray} 
and the Fadeev-Popov ghost term as
$$
\delta S_{FP}  =  \frac{\eta}{q}
\int r^2 dr \int d \Omega  \eta^a \left[ (D^2)_{ab} 
+ H^2 (\delta^{ab} -\hat{r}^a \hat{r}^b) \right] \eta^b . 
$$
where the key terms are given by 
\begin{eqnarray}
(D^2 )_{ab} & = & \delta_{ab}\partial_0^2 +2\epsilon_{abc}\hat{r}^cJ\partial_0 
-\delta_{ab}J^2+ \delta_{ab} \left( -\nabla_r^2 + \frac{L^2}{r^2} 
+ \frac{W^2}{r^2} \right)  - 2\frac{W}{r^2} \epsilon_{iab} L_i 
+ \hat{r}^a \hat{r}^b \frac{W^2}{r^2}  \nonumber \\
G_{ij}^c & = & \epsilon^{kij} \left[ \hat{r}^k \hat{r}^c 
\left( \frac{W^{\prime}}{r} + \frac{W}{r^2} (W-2) \right)  
- \delta^{kc}\frac{W^{\prime}}{r}  \right] \nonumber \\
(D_i \phi )^c & = &  \hat{r}^i \hat{r}^c\left( H^{\prime}-\frac{H}{r}(1-W)
\right) + \delta^{ic} \frac{H}{r} (1-W) \nonumber \\
G_{0i}^c & = & \hat{r}^i \hat{r}^c\left( J^{\prime}-\frac{J}{r}(1-W)
\right) + \delta^{ic} \frac{J}{r} (1-W) \nonumber \\
(D_0 \phi )^a & = & 0.  \nonumber
\end{eqnarray}
Note that in this case there is a linear
time derivative so it is not obvious that it is possible to set up a 
standard eigenvalue problem.

The dyon solution is also invariant under rotations generated by $K$, and so
we again expand the perturbations in eigenstates of $K$,
\begin{eqnarray}
\phi^a & = & \sum_{km} \sum_{j=-1}^{1} \phi^{km}_{j} Y^a_{kjm} \nonumber \\
\eta^a & = & \sum_{km} \sum_{j=-1}^{1} \eta^{km}_{j} Y^a_{kjm} \nonumber \\
W_0^a & = & \sum_{km} \sum_{j=-1}^{1} \tilde{W}^{km}_{j} Y^a_{kjm} \nonumber \\
W_i^a & = & \sum_{km} \sum_{j_1=0}^{2} \sum_{j_2=-j_1}^{j_1} 
W^{km}_{j_1j_2} Y^{ai}_{kj_1j_2m} \nonumber
\end{eqnarray}
and again find that the perturbations separate into two channels under parity
$$
\psi^+_{km} = \left( \begin{array}{c}
                W^{km}_{22} \\ W^{km}_{20} \\ W^{km}_{2-2} \\ W^{km}_{10} \\
                W^{km}_{00} \\ \tilde{W}^{km}_{0} \\ \phi^{km}_{0}
                \end{array} \right), \hspace{20mm}
\psi^-_{km} = \left( \begin{array}{c}
                W^{km}_{21} \\ W^{km}_{2-1} \\ W^{km}_{11} \\ W^{km}_{1-1} \\
                \tilde{W}^{km}_{1} \\ \tilde{W}^{km}_{-1} \\ 
                \phi^{km}_{1} \\ \phi^{km}_{-1}
                \end{array} \right).
$$
Substituting these expansions into $\delta S_{gf}$ and performing the 
angular integrals gives the perturbation operator 
\begin{eqnarray}
M_{\alpha \beta}^{(\pm)} & = & \delta_{\alpha \beta} \left[
\partial_0^2 -\nabla_r^2 + \frac{(k+j_2 )(k+j_2 +1)}{r^2} + \frac{W^2}{r^2} 
+ H^2 -J^2 +\frac{\beta}{2} (H^2 -1) P_{\phi} \right] \nonumber \\
 & + & 2\frac{W}{r^2} C_{\alpha\beta}^{(\pm )(1)}
 + 2J C_{\alpha\beta}^{(\pm )(9)} \partial_0
 + \left( \frac{W^2}{r^2} - H^2 +J^2 + \beta H^2 P_{\phi} \right)
 C_{\alpha\beta}^{(\pm )(2)} \nonumber \\
 & + & 2\frac{W^{\prime}}{r} C_{\alpha\beta}^{(\pm )(3)}
  +  2 \frac{W}{r^2} (W-2) C_{\alpha\beta}^{(\pm )(4)} \nonumber \\
 & + & 4\frac{H}{r} (1-W) C_{\alpha\beta}^{(\pm)(5)}
 + 4 \left( H^{\prime} - \frac{H}{r} (1-W) \right) 
 C_{\alpha\beta}^{(\pm )(6)} \nonumber  \\
 & + & 4\frac{J}{r} (1-W) C_{\alpha\beta}^{(\pm)(7)} \epsilon_{\alpha\beta}
 + 4 \left( J^{\prime} - \frac{J}{r} (1-W) \right) 
 C_{\alpha\beta}^{(\pm )(8)} \epsilon_{\alpha\beta} \nonumber 
\end{eqnarray}
for the Higgs and gauge fields, and
\begin{eqnarray}
\tilde{M}_{\alpha\beta}^{(\pm)} & = & \delta_{\alpha\beta}
\left[ -\nabla_r^2 + \frac{(k+j_2)(k+j_2+1)}{r^2} + \frac{W^2}{r^2} + H^2
 -J^2 \right] \nonumber \\
& +& \frac{2W}{r^2} C_{\alpha\beta}^{(\pm)(1)} 
+ 2J C_{\alpha\beta}^{(\pm )(9)} \partial_0
+ \left( \frac{W^2}{r^2} - H^2 +J^2 \right) C_{\alpha\beta}^{(\pm)(2)} 
\nonumber
\end{eqnarray}
for the ghosts. We have used the same notation as for the monopole and the 
non-zero elements of the matrices $C_{\alpha\beta}^{(\pm)(1-6)}$ are the same
as for the monopole. The non-zero elements of the matrices 
$C_{\alpha\beta}^{(\pm)(7,8)}$ are the same as those of the matrices
$C_{\alpha\beta}^{(\pm)(5,6)}$ but with the $\alpha,\beta$ refering to the
$\tilde{W}$ fields instead of the $\phi$ fields. We have not calculated the 
elements of $C_{\alpha\beta}^{(\pm)(9)}$ for general $k$, but for the case of
interest ($k=0$) the elements are all zero, and so in this case we can set up a 
standard eigenvalue problem as for the monopole.

We will restrict our attention to the Bogomol'nyi limit and consider the
analytic form of the dyon solution given earlier. Substituting this solution 
into the perturbation equation and rescaling
$$
x \rightarrow (1-C^2)^{-1/2} x \equiv (1+e^2)^{1/2} x, \hspace{5mm}
t \rightarrow (1-C^2)^{-1/2} t \equiv (1+e^2)^{1/2} t
$$
gives $M_{\alpha\beta}^{(\pm)} = (1+e^2)^{-1} 
\overline{M}_{\alpha\beta}^{(\pm)}$
where
\begin{eqnarray}
\overline{M}_{\alpha \beta}^{(\pm)} & = & \delta_{\alpha \beta} \left[
\partial_0^2 -\nabla_r^2 + \frac{(k+j_2 )(k+j_2 +1)}{r^2} + \frac{W^2}{r^2} 
+ H^2 -J^2 +\frac{\beta}{2} (H^2 -1) P_{\phi} \right] \nonumber \\
 & + & 2\frac{W}{r^2} C_{\alpha\beta}^{(\pm )(1)}
 + \left( \frac{W^2}{r^2} - H^2 +J^2 + \beta H^2 P_{\phi} \right)
 C_{\alpha\beta}^{(\pm )(2)} \nonumber \\
 & + & 2\frac{W^{\prime}}{r} C_{\alpha\beta}^{(\pm )(3)}
  +  2 \frac{W}{r^2} (W-2) C_{\alpha\beta}^{(\pm )(4)} \nonumber \\
 & + & 4\sqrt{1+e^2} \left( \frac{H}{r} (1-W) 
 C_{\alpha\beta}^{(\pm)(5)}+ \left( H^{\prime} - \frac{H}{r} (1-W) \right) 
 C_{\alpha\beta}^{(\pm )(6)} \right) \nonumber  \\
 & + & 4e \left( \frac{J}{r} (1-W) 
 C_{\alpha\beta}^{(\pm)(7)} \epsilon_{\alpha\beta} + \left( J^{\prime} 
 - \frac{J}{r} (1-W) \right) 
 C_{\alpha\beta}^{(\pm )(8)} \epsilon_{\alpha\beta} \right). \nonumber 
\end{eqnarray}
Note that the quantity we have used in the rescaling $(1-C^2)^{-1/2}$ is equal 
to
$\sqrt{1+e^2}$, which is the ratio of dyon mass to the monopole mass.

\subsection*{$k=0$ modes}

If we now consider the $k=0$ modes we find that the allowed modes are
$$
\psi_{00}^+ = \left( \begin{array}{c}
                     W_{22}^{00} \\  W_{00}^{00}
                     \end{array} \right)  \hspace{15mm}
\psi_{00}^- = \left( \begin{array}{c}
                     W_{11}^{00} \\ \tilde{W}_{1}^{00} \\ \phi_{1}^{00}
                     \end{array} \right). 
$$
$\psi_{00}^+$ satisfies the eigenvalue equation 
$$
\overline{M} \psi_{00}^+ = \omega^2 \psi_{00}^+
$$
where the elements of $\overline{M}$ are the same as for the monopole, and
so this equation is the same as equation \ref{+zero}. So we have the same
bound state solutions with eigenvalues
$$
\omega_n^2 = 1 - \frac{1}{n^2} \hspace{5mm} \mbox{where} \hspace{5mm}
n=1,2,3,\ldots
$$
as for the monopole. Note that there is explicitly no $e$-dependence of the
eigenvalues.

The equation for $\psi_{00}^-$ is different for the dyon and does possess
some $e$-dependence. $\psi_{00}^-$ satisfies the eigenvalue equation
$$
\overline{M} \psi_{00}^-  = \omega^2 \psi_{00}^-
$$
where
\begin{eqnarray}
M_{11} & = &  -\nabla_r^2 + \frac{2(1-W)^2}{r^2} + H^2 + \frac{W(W-2)}{r^2}
\nonumber \\
M_{22} & = & M_{33} \hspace{3mm} = \hspace{3mm} -\nabla_r^2 
+ \frac{2(1-W)^2}{r^2} \nonumber \\
M_{12} & = & -M_{21} \hspace{3mm} = \hspace{3mm} 
 e \: 2\sqrt{2} \frac{H(1-W)}{r} \nonumber \\
M_{13} & = & M_{31} \hspace{3mm} = \hspace{3mm} \sqrt{1+e^2}\: 2\sqrt{2} 
\frac{H(1-W)}{r} \nonumber \\
M_{23} & = & M_{32} \hspace{3mm} = \hspace{3mm} 0 \nonumber
\end{eqnarray}
However in the far field limit the perturbation equation becomes
$$
\left( \begin{array}{ccc}
   -\nabla_r^2 + 1 -\frac{2}{r} & 0 & \\
   0 & -\nabla_r^2 & \\
    0 & 0 & -\nabla_r^2
    \end{array} \right)
                    \left( \begin{array}{c}
                     W_{11}^{00} \\ \tilde{W}_{1}^{00} \\ \phi_{1}^{00}
                     \end{array} \right) 
                     = \omega^2 \left( \begin{array}{c}
                     W_{11}^{00} \\ \tilde{W}_{1}^{00} \\ \phi_{1}^{00}
                     \end{array} \right)
$$
which has no $e$-dependence. So as for the monopole we find that the 
$W_{11}^{00}$ field has bound solutions with eigenvalues
$$
\omega_n^2 = 1 - \frac{1}{n^2} \hspace{5mm} \mbox{where} \hspace{5mm}
n=1,2,3,\ldots
$$
The $k=0$ ghost equation for the dyon is the same as for the monopole, and
so in the far field limit the ghosts cancel the $\tilde{W}_{1}^{00}$ and 
$\phi_{1}^{00}$ massless modes just as for the monopole.

If we were to solve the $\psi_{00}^-$ eigenvalue equation with the full
analytic solution, then from our results for the monopole we would expect
the eigenvalues to agree with the analytic far field result i.e. with no
$e$-dependence. On physical grounds, however, we would expect some of the 
eigenvalue spectra to possess an $e$-dependence because there should be an
interaction between the electric charge of the dyon and the electric charge
of the $W$-boson. The situation is unfortunately obscured by the fact that
eigenstates of $K$ are not eigenstates of charge. The perturbation equations
for the higher $k$ modes do in general possess an $e$-dependence in the far 
field limit, but we have not solved the perturbation equations in these 
cases.

\section{Perturbations about the point electric solution}
 
Substituting the point electric solution into the perturbation equations 
\ref{pert} gives, in the Bogomol'nyi limit
\begin{equation} \label{massless}
\left( \begin{array}{ccc}
 \overline{D}^2 & 0 & 0 \\ 0 & \overline{D}^2 & 0 \\ 
 0 & 0 & \overline{D}^2
 \end{array} \right) \left( \begin{array}{c}
                \phi^3 \\ W_0^3 \\ W_i^3
                \end{array} \right) = 0,
\end{equation}
and
\begin{equation} \label{massive}
\left( \begin{array}{ccc}
 D^2 +(\phi^3)^2 & 0 & \pm 2iD_i\phi^3 \\
 0 & D^2 +(\phi^3)^2 & \mp 2iG_{0i}^3 \\
 \mp 2iD_i\phi^3 & \mp 2iG_{0i}^3 & D^2 +(\phi^3)^2
 \end{array} \right) \left( \begin{array}{c}
                \phi^{\pm} \\ W_0^{\pm} \\ W_i^{\pm}
                \end{array} \right) = 0
\end{equation}
where
\begin{eqnarray}
\overline{D}^2 & = & -\omega^2 - \nabla_r^2 + \frac{L^2}{r^2} \nonumber \\
D^2 & = & -\left( \omega \mp \frac{Z}{r} \right)^2 - \nabla_r^2 
+ \frac{L^2}{r^2} \nonumber \\
D_i\phi^3 & = & Z\frac{\hat{r}_i}{r^2} \nonumber \\
G_{0i}^3 & = & Z\frac{\hat{r}_i}{r^2} .\nonumber
\end{eqnarray}
The equations for the gosts $(\eta^3,\eta^{\pm})$ are
\begin{eqnarray}
\overline{D}^2 \eta^3 & = & 0 \nonumber \\
\left( D^2 +(\phi^3)^2 \right) \eta^{\pm} & = & 0. \nonumber
\end{eqnarray}
The point electric solution is invariant under the total angular momentum 
operator $J=L+S$, and so we expand the fields in terms of spherical harmonics
$Y_{jm}$ and vector spherical harmonics $Y^a_{jsm}$ as
\begin{eqnarray}
\phi & = & \sum_{jm} \phi_{jm} Y_{jm} \nonumber \\
W_0 & = & \sum_{jm} \tilde{W}_{jm} Y_{jm} \nonumber \\
\eta & = & \sum_{jm} \eta_{jm} Y_{jm} \nonumber \\
W_i & = & \sum_{jm} \sum_{s=-1}^1 W_s^{jm} Y^i_{jsm} \nonumber
\end{eqnarray}
where we have suppressed the $\pm,3$ isospin labels. The angular momentum $l$ 
is given by $l=j+s$.

Unlike the cases of the monopole and dyon, in this case because the total
angular momentum operator is independent of the isospin generator $T$ we can 
construct  simultaneous eigenstates of $J$ and the charge operator $Q=T_3$. 
The $(\phi^3,W^3_{\mu})$ perturbations therefore have zero charge and so the 
physical modes of equation \ref{massless} involve the neutral massless gauge 
boson and the neutral Higgs scalar. The $(\phi^{\pm},W^{\pm}_{\mu})$ 
perturbations have charges $\pm 1$ and so the physical modes of equation 
\ref{massive} involve the massive charged gauge bosons.

\subsection*{$j=0$ modes}

If we consider the $j=0$ modes then we find that the only allowed mode is
$$
\left( \begin{array}{c}
                 \phi^{00} \\ \tilde{W}^{00} \\ W^{00}_{1}
                 \end{array} \right),
$$
which satisfies the perturbation equation
\begin{equation} \label{epert}
\left( \begin{array}{c}
-\left(\omega -\frac{Z}{r}\right)^2 -\nabla_r^2 +\left(1-\frac{Z}{r} \right)^2
\hspace{13mm} 0 \hspace{13mm} 2iZ/r^2 \\
0 \hspace{11mm} 
-\left(\omega -\frac{Z}{r}\right)^2 -\nabla_r^2 +\left(1-\frac{Z}{r} \right)^2
\hspace{11mm} -2iZ/r^2 \\
-2iZ/r^2 \hspace{3mm} -2iZ/r^2 \hspace{3mm}
-\left(\omega -\frac{Z}{r}\right)^2 -\nabla_r^2 + \frac{2}{r^2}
+\left(1-\frac{Z}{r} \right)^2
  \end{array} \right) \left( \begin{array}{c}
                       \phi^{00} \\ \tilde{W}^{00} \\ W^{00}_{1}
                 \end{array} \right)   = 0,
\end{equation}
where we have taken the matrix elements for $(\phi^+,W_{\mu}^+)$ in equation 
\ref{massive}. This is not of the form of a standard eigenvalue problem as it
stands, but if we rescale $\rho =2(1-\omega^2)^{1/2} r$ and define
$$
\Lambda = Z \frac{(1-\omega)}{\sqrt{1-\omega^2}}
$$
then we obtain
\begin{equation} 
\left( \begin{array}{c}
-\nabla_{\rho}^2 +\frac{1}{4}-\frac{\Lambda}{\rho}
\hspace{14mm} 0 \hspace{14mm} 2iZ/\rho^2 \\
0 \hspace{11mm} -\nabla_{\rho}^2 +\frac{1}{4}-\frac{\Lambda}{\rho}
\hspace{11mm} -2iZ/\rho^2 \\
-2iZ/\rho^2 \hspace{3mm} -2iZ/\rho^2 \hspace{3mm}
-\nabla_{\rho}^2 +\frac{1}{4}-\frac{\Lambda}{\rho} +\frac{2}{\rho^2}
  \end{array} \right) \left( \begin{array}{c}
                       \phi^{00} \\ \tilde{W}^{00} \\ W^{00}_{1}
                 \end{array} \right)   = 0,
\end{equation}
which is now an eigenvalue problem with eigenvalues $\Lambda$. If we rewrite 
this as
$$
\left[ \left( \nabla_{\rho}^2 -\frac{1}{4}+\frac{\Lambda}{\rho} \right) 
\underline{I}_3 +\frac{2}{\rho^2} \underline{A} \right] \left(
\begin{array}{c}
  \phi^{00} \\ \tilde{W}^{00} \\ iW^{00}_{1}
    \end{array} \right)   = 0,
$$
where $\underline{I}_3$ is the $3\times 3$ identity and the matrix 
$\underline{A}$ is given by
$$
\underline{A} = \left( \begin{array}{ccc}
  0 & 0 & -Z \\ 0 & 0 & Z \\ -Z & -Z & -1
  \end{array} \right),
$$
then the problem of diagonalizing the perturbation matrix has been reduced to
diagonalizing the matrix $\underline{A}$, which is easy. The eigensolutions
of 
$$
\underline{A}\: e = \Omega \: e
$$
are
$$
\Omega_1=0 \hspace{5mm} e_1=\left( 
\begin{array}{c} 1 \\ 1 \\ 0 \end{array} \right), \hspace{5mm}
\Omega_2=0 \hspace{5mm} e_2=\left( 
\begin{array}{c} 1 \\ -1 \\ 0 \end{array} \right), \hspace{5mm}
\Omega_3=-1 \hspace{5mm} e_3=\left(
\begin{array}{c} Z \\ -Z \\ 1 \end{array} \right) .
$$
The corresponding eigenvectors of the perturbation matrix are
$$
\left( \begin{array}{c}
\phi^{00} \\ \tilde{W}^{00} \\ iW^{00}_{1} \end{array} \right)
\hspace{3mm} =  \hspace{3mm}
\left( \begin{array}{c} 1 \\ 1 \\ 0 \end{array} \right) w_1(\rho),\hspace{5mm}
\left( \begin{array}{c} 1 \\ -1 \\ 0 \end{array} \right) w_2(\rho),\hspace{5mm}
\left( \begin{array}{c} Z \\ -Z \\ 1 \end{array} \right) w_3(\rho),
$$
where the functions $w_i(\rho)$ satisfy
\begin{eqnarray}
\left( \nabla_{\rho}^2 -\frac{1}{4}+\frac{\Lambda}{\rho} \right) w_i(\rho) 
& = & 0 \hspace{5mm} \mbox{for} \hspace{3mm} i=1,2 \nonumber \\
\left( \nabla_{\rho}^2 -\frac{1}{4}+\frac{\Lambda}{\rho} -\frac{2}{\rho^2} 
\right) w_3(\rho) & = & 0. \nonumber
\end{eqnarray}
If we write $w_i(\rho) =u_i(\rho)/\rho$, then the equations for the $u_i$ are of 
the form of Whittakers equation
$$
\left( \frac{d^2}{d\rho^2} -\frac{1}{4} + \frac{\Lambda}{\rho}
+\frac{\frac{1}{4} -\mu^2}{\rho^2} \right) u_i(\rho )=0.
$$
where $\mu =1/2$ for $u_1$ and $u_2$, and $\mu =3/2$ for $u_3$. As we stated
earlier, for $\omega^2 \geq 1$ (imaginary $\rho$) these equations have Whittaker
function solutions $W_{\Lambda,\mu}(\rho)$ for all $\Lambda$ which are 
oscillatory
at large $\rho$. For $1 > \omega^2 \geq 0 $ (real $\rho$) the Whittaker fuction 
solutions only exist for
$$
\Lambda = n_r + \mu + \frac{1}{2} \hspace{10mm} n_r = 0,1,2,3,\ldots
$$
and the solutions decay exponentially at large $\rho$.

At this point it is worth considering the ghosts.
The corresponding equation for the $j=0$ mode of the charged ghosts is
\begin{equation} \label{eghosts}
\left[ -\left(\omega -\frac{Z}{r}\right)^2 -\nabla_r^2 
+\left(1-\frac{Z}{r} \right)^2 \right] \eta_{00} =0.
\end{equation}
If we similarly rescale $\rho = 2(1-\omega^2)^{1/2} r$ and define 
$$
\Lambda = Z \frac{(1-\omega )}{\sqrt{1-\omega^2}}
$$
then we can rewrite the ghost equation as
$$
\left( \nabla_{\rho}^2 - \frac{1}{4} + \frac{\Lambda}{\rho} \right) \eta_{00}
=0,
$$
which is the same as the equation for the $w_1$ and $w_2$ eigenmodes. The ghosts
therefore cancel the $w_1$ and $w_2$ eigenmodes leaving the $w_3$ eigenmode as
the physical mode.

For the $w_3$ mode we have $\mu=3/2$ and so there are physical bound modes for
$$
\Lambda = n = 2,3,4,\ldots
$$
Substituting this into the expression
$$
\Lambda = Z \frac{(1-\omega )}{\sqrt{1-\omega^2}}
$$
and squaring it gives a quadratic equation in $\omega$ with the two solutions
$$
\omega_+=1, \hspace{5mm} \mbox{and} \hspace{5mm}
\omega_- = -\frac{(1 - Z^2/n^2)}{(1 + Z^2/n^2)} \hspace{10mm} n=2,3,4,\ldots
$$
For our point electric solution to correspond to a $W^+$ boson we take 
$Z=1/4\pi $ in our rescaled units. Since then $Z^2 =1/16\pi^2 < 1$ we can
approximate $\omega_-$ as
$$
\omega_- \simeq -\left( 1-\frac{2Z^2}{n^2} \right) \hspace{5mm} n=2,3,4,\ldots
$$
Now the perturbations are eigenstates of charge and so we have
$$
W_{\mu}^{\pm} (\underline{x},t)=W_{\mu}^{\pm} (\underline{x}) 
{\rm e}^{\pm i\omega t}.
$$
So the $\omega_+$ solution therefore corresponds to the $W^+$ boson
and the $\omega_-$ solution corresponds to the $W^-$ boson.
Our result therefore agrees with the result of ref.~\cite{Manton77} where
it is shown that the force between two gauge bosons of the same charge is
zero, and hence no bound states, whilst the force between gauge bosons
of opposite charges is twice the electrostatic coulomb force.

We note that the point electric solution used here has no perturbative
instabilities ($\omega$ imaginary), no matter how large the charge $4\pi Z$.
This differs from the point electric solution
$$
W_0^3 = - \frac{Z}{r}, \hspace{5mm} \phi^3 = \eta
$$
which possesses perturbative instabilities above a critical charge 
\cite{Mandula77}.

\section{Discussion}

We have found the eigenvalue spectra for spherically symmetric perturbations
about the monopole, dyon and point electric solutions. For the perturbative
modes of positive parity about both the monopole and dyon we argue that the
eigenvalue spectra of bound modes are the same using both the far field form
and the full form of the Prasad-Sommerfield solution. These bound states are
interpreted as bound states of the massive $W$-bosons in the presence of the 
monopole and the dyon. The force between the monopole and the massive gauge
bosons consists of two components. One part is the electromagnetic interaction 
between the monopole magnetic charge and the magnetic dipole of the gauge 
boson. However it was shown in ref.~\cite{Olsen90} that this interaction by 
itself is not strong enough to give bound states. The other part of the 
monopole-gauge boson force arises from the Higgs field of the monopole. 
Because the gauge boson acquires its mass from the Higgs field expectation 
value, the gauge boson has a lower energy when it is in the presence of the 
monopole core and this gives rise to a force which is responsible for confining 
the gauge boson to the core of the monopole.
The force between the dyon and the massive gauge bosons has an electrostatic
force in addition to the monopole-gauge boson force. The effect of this
electrostatic force is unfortunately obscured by the fact that
the perturbative modes about the dyon (and monopole) are not in eigenstates of
charge. This is not due to any failings of our perturbative method but is an
intrinsic property of both the dyon and monopole solutions.

The perturbations about the point electric solution are in eigenstates of
charge. The force between two charged gauge bosons consists of an attractive
Higgs force and an electrostatic force of equal magnitude. For gauge bosons of
equal charge the two forces cancel and in this case we find no bound states,
whereas for oppositely charged bosons the two forces add and our results for
the bound states in this case are consistent with this.

We set out to answer the question: are the bound state energies of the various
two particle systems of the Yang-Mills-Higgs model related by electromagnetic
duality? We are specifically considering the $SL(2,{\cal Z})$ symmetry of Sen
\cite{Sen94}. The original conjecture of Montonen and Olive \cite{Montonen77}
was that there are two equivalent choices of action, either with the 
$W$-bosons as the gauge particles and the monopoles as the solitons, or with
the monopoles as the gauge particles and the $W$-bosons as the solitons. The
spectrum of dyons does not, however, fit nicely into this picture. In the 
electromagnetic duality of Sen \cite{Sen94} the electric $(q)$ and magnetic 
$(g)$ charges of a particle state are given by 
$$
q+ig = q_0 (m\tau + n) \hspace{10mm} m,n \in {\cal Z}
$$
where $q_0$ is the minimum electric charge and
$$
\tau = \frac{\theta}{2\pi} + \frac{i}{q_0^2}.
$$
$\theta$ is a new parameter of the theory. Plotting $q+ig$ in the complex
plane gives the charge lattice $(q,g)$, and the single particle states 
correspond to the primitive vectors of this charge lattice. A basis for the 
charge lattice is given by any two non-collinear primitive vectors, and choices 
of basis correspond to choices of equivalent actions. Invariance of the theory 
under electromagnetic duality is now realised as invariance of the theory 
to changes of basis of the charge lattice.

If we choose the primitive vector $q_0$ to correspond to the electric gauge 
boson and the primitive vector $q_0\tau $ to correspond to the magnetic
monopole, then a change of basis to $(q_0^{\prime},q_0^{\prime}\tau^{\prime})$
is given by \cite{Olive95}
$$
\left( \begin{array}{c}
 q_0^{\prime}\tau^{\prime} \\ q_0^{\prime}
 \end{array} \right) = A \left( \begin{array}{c}
 q_0\tau \\ q_0
 \end{array} \right) = \left( \begin{array}{cc}
              a & b \\ c & d
              \end{array} \right)  \left( \begin{array}{c}
                                 q_0\tau \\ q_0
                                  \end{array} \right), 
$$
where $a,b,c,d \in {\cal Z}$ and $ad - bc =1$, so the matrices $A$ form the
group $SL(2,{\cal Z})$. Under this transformation $\tau$ becomes
$$
\tau^{\prime} = \frac{a\tau + b}{c\tau +d}.
$$
We now consider whether our results are consistent with this picture. Our
$W$-boson-monopole system corresponds to choosing the primitive vectors
$q_0$ as the $W$-boson and $q_0\tau=i/q_0$ as the monopole. In this case we
find that the spherically symmetric bound perturbative modes about the monopole
have energies
$$
\omega^2_n = \left(1-\frac{1}{n^2}\right) 
\equiv \left(1-\frac{qg}{n^2}\right) \hspace{5mm}
n \in {\cal Z}
$$
where $q$ is the charge of the $W$-boson and $g$ is the charge of the monopole.
If we now apply the $SL(2,{\cal Z})$ transformation
\begin{eqnarray}
\left( \begin{array}{c}
 q_0^{\prime}\tau^{\prime} \\ q_0^{\prime}
 \end{array} \right) & = & \left( \begin{array}{cc}
              1 & 1 \\ 0 & 1
              \end{array} \right)  \left( \begin{array}{c}
                                 q_0\tau \\ q_0
                                  \end{array} \right) =
                                 \left( \begin{array}{c}
                                 q_0\tau + q_0 \\ q_0
                                  \end{array} \right) \nonumber \\
\tau^{\prime} & = & \tau +1 \nonumber
\end{eqnarray}
then the $W$-boson still corresponds to 
the point $q_0$, but the monopole in the new basis corresponds to the point 
$q_0\tau + q_0$, which is the dyon of the old basis. The product of the
$W$-boson charge $q$ and the monopole charge $g$ is invariant under this 
transformation, so the bound state energies of the two particle system should
be invariant under this transformation. We did indeed find that the 
$W$-boson-dyon bound state energies are given by the same expression, despite
our reservations about the $W$-field perturbations not being in eigenstates of
charge.

The $SL(2,{\cal Z})$ transformation
\begin{eqnarray}
\left( \begin{array}{c}
 q_0^{\prime}\tau^{\prime} \\ q_0^{\prime}
 \end{array} \right) & = & \left( \begin{array}{cc}
              0 & -1 \\ 1 & 0
              \end{array} \right)  \left( \begin{array}{c}
                                 q_0\tau \\ q_0
                                  \end{array} \right) =
                                 \left( \begin{array}{c}
                                  - q_0 \\ q_0\tau
                                  \end{array} \right)  \nonumber \\
\tau^{\prime} & = & -\frac{1}{\tau} \nonumber                                  
\end{eqnarray}
applied to the original $W$-boson-monopole basis $(q_0,i/q_0)$ corresponds
to the duality transformation of Montonen and Olive \cite{Montonen77}. The
transformation displays the interchange of weak and strong coupling
$\tau \rightarrow -1/\tau$.
This weak-strong coupling interchange also occurs for the transformation
\begin{eqnarray}
\left( \begin{array}{c}
 q_0^{\prime}\tau^{\prime} \\ q_0^{\prime}
 \end{array} \right) & = & \left( \begin{array}{cc}
              1 & -1 \\ 1 & 0
              \end{array} \right)  \left( \begin{array}{c}
                                 q_0\tau \\ q_0
                                  \end{array} \right) =
                                 \left( \begin{array}{c}
                                  q_0\tau- q_0 \\ q_0\tau
                                  \end{array} \right)  \nonumber \\
\tau^{\prime} & = & 1 -\frac{1}{\tau}, \nonumber                                  
\end{eqnarray}
where the new gauge boson again corresponds to the old monopole but the new
monopole corresponds to the old dyon with charges $(-q_0,i/q_0)$ in the
original basis. The product $qg$ is again invariant under this transformation
and so we would expect the monopole-dyon bound state energies to again be
given by
$$
\omega^2_n = \left(1-\frac{1}{n^2}\right) 
\equiv \left(1-\frac{qg}{n^2}\right).
$$
The monopole-dyon two particle system was studied in ref.~\cite{Gibbons86}
where they found physical bound states for
\begin{eqnarray}
\epsilon & = & \frac{1}{2} (n^2-s^2)^{1/2} (n-(n^2-s^2)^{1/2}) \nonumber \\
    & \simeq & \frac{s^2}{4} \left( 1-\frac{s^2}{4n^2} \right) \hspace{5mm}
    \mbox{for} \hspace{5mm} n^2>s^2 ,\nonumber
\end{eqnarray}
where $s,n$ are integers and $s$ is the relative electric charge between the 
monopole and dyon. This is obviously different from what we have above.

Both our results and those of ref.~\cite{Gibbons86} are only approximations,
and under the $SL(2,{\cal Z})$ duality transformations they correspond to 
different limits. Our approach was perturbative, and we found the harmonic
oscillator states about the monopole and dyon. Although in the quantum theory
these quantized harmonic oscillator states are interpreted as the gauge bosons 
of the theory, this approach only accurately describes the $W$-boson-soliton 
two particle system in the limit $q^2 \ll 1$.

In ref.~\cite{Gibbons86} they considered the dynamics of two non-relativistic
BPS monopoles using the Atiyah-Hitchin metric on the space of collective
co-ordinates of the monopoles. These collective co-ordinates consist of the
three spatial positions and a phase. The spherically symmetric zero mode
we obtained for both the monopole and dyon corresponds to changes in this 
phase. In the limit of large monopole separation they used the Taub-NUT
metric as an approximation to the Atiyah-Hitchin metric, and solved the
non-relativistic Schr\"odinger equation in Taub-NUT space for the relative
motion of the two monopoles.

The result of ref.~\cite{Gibbons86} is non-relativistic and was obtained in 
the limit of weak electric coupling. If we consider the energy of the dyon 
given earlier in the limit $q^2 \ll 1$ we find
$$
E_d =  \frac{4\pi\eta}{q} \sqrt{1+\left(\frac{Qq}{4\pi}\right)^2}
\simeq  \frac{4\pi\eta}{q} \left( 1+\frac{1}{2} 
\left(\frac{Qq}{4\pi}\right)^2 \right).
$$
So the total energy of the static monopole-dyon system in the limit of 
infinite separation for $q^2 \ll 1$ is
$$
E=E_d+E_m \simeq 4\pi \frac{\eta}{q}  \left( 2+\frac{1}{2} 
\left(\frac{Qq}{4\pi}\right)^2 \right).
$$
If we explicitly include factors of $\hbar$ so that the gauge coupling is 
$\hbar q$, then the weak coupling condition is $\hbar q \ll 1$. Now we have 
taken $\hbar =1$, and so weak coupling condition is $q \ll 1$, whereas in 
ref.~\cite{Gibbons86} they use units $q=1$ and so the weak coupling condition 
is given by $\hbar \ll 1$. With the units $q=1$, $\eta=1$ of 
ref.~\cite{Gibbons86} the energy given above becomes
$$
E \simeq 2 \left( 4\pi + \frac{Q^2}{16\pi} \right),
$$
and in the centre of motion frame with the mass and length in units of $4\pi$,
bound states occur for 
$$
0 < \epsilon < \frac{Q^2}{4}.
$$
Now the electric charge of a dyon in the presence of a monopole is not
a good quantum number, but the relative electric charge $(s)$ is. If, in the
centre of motion frame, we assign electric charges $(Q_1=-s/2)$ to the
monopole and $(Q_2=s/2)$ to the dyon, then the result
$$
\epsilon \simeq \frac{s^2}{4} \left( 1-\frac{s^2}{4n^2} \right)
$$
is consistent with an electrostatic interaction of the form $Q_1Q_2/r$
between the two particles \cite{Manton85}.

So both here and in ref.~\cite{Gibbons86} bound states due to a Coulomb
interaction are found in the limit $q^2 \ll 1$. However, the 
$SL(2,{\cal Z})$ transformation between the $W$-boson-monopole and 
monopole-dyon systems involves a weak-strong coupling interchange. So our
result for the $W$-boson-monopole system, which holds for $q^2 \ll 1$,
would correspond to the limit $q^2 \gg 1$ for the monopole-dyon system,
and the result of ref.~\cite{Gibbons86}, which holds for $q^2 \ll 1$ for the
non-relativistic monopole-dyon system, would correspond to the limit
$q^2 \gg 1$ for the non-relativistic $W$-boson-monopole system.

In ref.~\cite{Manton85} Manton showed that the monopole-dyon interaction
could be understood in terms of the motion of classical point particles with
electric and magnetic charges. If we consider a dyon with electric charge $q$
and magnetic charge $g$, then the scalar charge must be $(g^2+q^2)^{1/2}$
\cite{Manton85}. So the Coulomb interaction between two static particles
with charges $(q_1,g_1)$ and $(q_2,g_2)$ is \cite{Manton85}
$$
V = \left[ g_1g_2 + q_1q_2 - (g_1^2+q_1^2)^{1/2}(g_2^2+q_2^2)^{1/2} \right]
\frac{1}{r}.
$$
Now if we consider a monopole $(0,g)$ and a $W$-boson $(q,0)$ we find
$$
V = -\frac{qg}{r},
$$
and for a dyon $(q,g)$ and a $W$-boson $(q,0)$ we find
\begin{eqnarray}
V & = & \left[ q^2 - q(g^2+q^2)^{1/2} \right] \frac{1}{r} \nonumber \\
  & \simeq & -\frac{qg}{r} \hspace{5mm} \mbox{for} \hspace{5mm} 
  g \gg q ,\nonumber
\end{eqnarray}
which is the same as the $W$-boson-monopole interaction for $q^2 \ll 1$
(note $g=1/q$). The monopole-dyon interaction is given by
\begin{eqnarray}
V & = & \left[ g^2 - g(g^2+q^2)^{1/2} \right] \frac{1}{r} \nonumber \\
  & \simeq & -\frac{q^2}{2r} \hspace{5mm} \mbox{for} \hspace{5mm} 
  g \gg q ,\nonumber
\end{eqnarray}
which gives the result for $n^2 \gg s^2$ of ref.~\cite{Gibbons86}. 
For $q \gg g$ we find
$$
V \simeq -\frac{qg}{r},
$$
which is the same as for the $W$-boson-monopole system. 
This simple argument supports the view that our result for the bound state
energy
$$
\omega_n^2 = \left( 1-\frac{qg}{n^2} \right),
$$
applies for $q \ll 1$ for the relativistic $W$-boson-monopole and $W$-boson-dyon
systems, and also for the relativistic monopole-dyon system for $q^2 \gg 1$.

If the bound state energies of these three two-particle systems are indeed 
related 
by $SL(2,{\cal Z})$ transformations on the charge lattice, then the result is in 
fact far more general. This is because any two non-collinear primitive vectors 
of
the charge lattice can be transformed to the basis $(q_0,i/q_0)$. Our result 
then
gives the relativistic bound state energies in the weak coupling limit 
$(q^2 \ll 1)$, and the result of ref.~\cite{Gibbons86} gives the 
non-relativistic
bound state energies in the strong coupling limit ($q^2 \gg 1$).

If we consider two general single particle states with the same charges $(g,q)$,
they exert no static Coulomb forces on each other since
$$
V = \left[ g^2 + q^2 - (g^2 +q^2)^{1/2}(g^2 +q^2)^{1/2} \right] \frac{1}{r} =0,
$$
and so they form no bound states with one another. Two general single particle
states with opposite charges, however, have a static Coulomb force
$$
V = \left[ -g^2 - q^2 - (g^2 +q^2)^{1/2}(g^2 +q^2)^{1/2} \right] \frac{1}{r} 
= -2 \frac{g^2 +q^2}{r},
$$
and can form bound states. By an $SL(2,{\cal Z})$ transformation we can 
transform this two particle system to the $W^+$--$W^-$ boson system, which we
found to have bound state energies
$$
\omega = -\frac{(1-q^2/n^2)}{(1+q^2/n^2)} \hspace{5mm} \mbox{for} \hspace{5mm}
q \ll 1.
$$

As to our original question of whether the bound states of the various two
particle systems are related by electromagnetic duality, we have no definitive
answer but our results would appear to be consistent with it.

Finally we note the possible existence of an unexpected coincidence. Our
numerical solution of the perturbations about the monopole gave the lowest
massive bound mode to have an eigenvalue of $\omega^2 \simeq 0.768\; m^2$ in the
Bogomol'nyi limit. We, however, argued that the omission of the origin in our
numerical method is responsible for it being different from the analytical
far field result of $\omega^2 = 0.75\; m^2$. In ref.~\cite{Goodband95a} we
numerically solved for the perturbations about the local string, and in the
Bogomol'nyi limit ($m_H = m_W$) we found a bound boson with eigenvalue
$\omega^2 \simeq 0.775\;m^2$. Our numerical method in this case also 
necessitates
the omission of th origin due to a $1/r^2$ centrifugal term. In addition the
string profiles had to be solved for numerically because there is no analytic
solution. The fields of this bound mode vary in exactly the same place that the
string profiles vary most, and so the value of this eigenvalue is exceptionally
sensitive to the accuracy of the string profiles. So we argue that our numerical 
result of $\omega^2 \simeq 0.775\; m^2$ would be consistent with the true 
eigenvalue being $\omega^2 = 0.75\; m^2$. If this is the case then the kink, 
string and monopole all possess bound bosons with an eigenvalue of 
$\omega^2 = 0.75\; m^2$ when they are in their respective Bogomol'nyi limits 
\cite{Bogomolnyi76} ($\beta = m_H^2/m_W^2 \rightarrow 0$ for the monopole; 
$\beta = m_H^2/m_W^2 \rightarrow 1$ for the local string;
and all values of the Higgs self coupling for the kink). It is precisely when
these theories are in their respective Bogomol'nyi limits that supersymmetric
extensions can be constructed where the Bogomol'nyi bounds \cite{Bogomolnyi76}
can be derived algebraically from the supersymmetry algebra. So it may be that
this apparently common bound of $\omega^2 \geq 0.75\;m^2$ for the bosons of the
theory in the presence of a defect with topological charge one, could be 
similarly derived in the supersymmetric extensions of the respective theories.

\subsection*{Acknowlodgements}
We are grateful to Mark Hindmarsh for useful discussions.
This work was supported by PPARC: studentship number 93300941.

\newpage

\section*{Table captions}

{\bf Table 1:} Diagonal elements of $C_{\alpha \beta }^{(\pm )(1-4)}$
\newline
{\bf Table 2:} Off-diagonal elements of $C_{\alpha \beta }^{(\pm )(1)}$ 
\newline
{\bf Table 3:} Off-diagonal elements of $C_{\alpha \beta }^{(\pm )(2-4)}$
\newline
{\bf Table 4:} Elements of $C_{\alpha \beta }^{(\pm )(5-6)}$
\newline
{\bf Table 5:} Bound mode eigenvalues for the monopole

\newpage

\begin{table}[t]
\centering
\begin{tabular}{|c|c|c|c|c|}  \hline
$\alpha =(j_1,j_2)$ & $C_{\alpha \alpha }^{(+)(1)}$ & 
$C_{\alpha \alpha }^{(+)(2)}$ & $C_{\alpha \alpha }^{(+)(3)}$ &
$C_{\alpha \alpha }^{(+)(4)}$ \\
\hline 
2,$-$2 & $k-2$ & $\frac{k-1}{2k-1}$ & $\frac{2(k-1)}{2k-1}$ & $\frac{-1}{2k-1}$ 
\\
2,0 & $-\frac{3}{2}$ & $\frac{9+4k(k+1)}{6(2k-1)(2k+3)}$ &
      $\frac{9+4k(k+1)}{3(2k-1)(2k+3)}$ & 
      $\frac{2(9-4k(k+1))}{3(2k-1)(2k+3)}$ \\
2,2 & $-(k+3)$ & $\frac{k+2}{2k+3}$ & $\frac{2(k+2)}{2k+3}$ & $\frac{1}{2k+3}$ 
\\
1,0 & $-\frac{1}{2}$ & $\frac{1}{2}$ & $-1$ & 0 \\
0,0 & 0 & $\frac{1}{3}$ & $-\frac{4}{3}$ & $\frac{2}{3}$ \\
0 & $-1$ & 0 & 0 & 0 \\
\hline \hline
$\alpha =(j_1,j_2)$ & $C_{\alpha \alpha }^{(-)(1)}$ & 
$C_{\alpha \alpha }^{(-)(2)}$ & $C_{\alpha \alpha }^{(-)(3)}$ &
$C_{\alpha \alpha }^{(-)(4)}$ \\
\hline 
2,$-$1 & $\frac{1}{2}(k-3)$ & $\frac{k-1}{2(2k+1)}$ & $ \frac{k-1}{2k+1}$ & 
      $-\frac{(k+2)}{2k+1}$ \\
2,1 & $-\frac{1}{2}(k+4)$ & $\frac{k+2}{2(2k+1)}$ & $\frac{k+2}{2k+1}$ &
       $-\frac{(k-1)}{2k+1}$ \\
1,$-$1 & $\frac{1}{2}(k-1)$ & $\frac{k+1}{2(2k+1)}$ & $-\frac{(k+1)}{2k+1}$ &
      $\frac{k}{2k+1}$ \\
1,1 & $-\frac{1}{2}(k+2)$ & $\frac{k}{2(2k+1)}$ & $-\frac{k}{2k+1}$ &
       $\frac{k+1}{2k+1}$ \\
$-1$ & $(k-1)$ & $\frac{k}{2k+1}$ & 0 & 0 \\
1 & $-(k+2)$ & $\frac{k+1}{2k+1}$ & 0 & 0 \\
\hline  
\end{tabular}
\caption{}
\end{table} 

\begin{table}[b]
\centering
\begin{tabular}{|c|c|c|}  \hline
$\alpha =(j_1,j_2)$ & $\beta =(j_1,j_2)$ & $C_{\alpha \beta }^{(+)(1)}$ \\
\hline  
1,0 & 0,0 & $-\left[ \frac{2}{3} k(k+1) \right]^{1/2}$ \\
2,0 & 1,0 & $\frac{1}{2} \left[ \frac{1}{3} (2k-1)(2k+3) \right]^{1/2}$ \\
\hline \hline
$\alpha =(j_1,j_2)$ & $\beta =(j_1,j_2)$ & $C_{\alpha \beta }^{(-)(1)}$ \\
\hline 
2,$-$1 & 1,$-$1 & $\frac{1}{2} \left[ (k-1)(k+1) \right]^{1/2}$ \\
2,1 & 1,1 & $\frac{1}{2} \left[ k(k+2) \right]^{1/2}$ \\
\hline
\end{tabular}
\caption{}
\end{table}

\begin{table}[t]
\centering
\begin{tabular}{|c|c|c|c|}  \hline
$\alpha =(j_1,j_2)$ & $\beta =(j_1,j_2)$ & 
$C_{\alpha \alpha }^{(+)(2)}$ & $C_{\alpha \alpha }^{(+)(3)}$=
$C_{\alpha \alpha }^{(+)(4)}$ \\
\hline
2,$-$2 & 2,0 &$-\left[ \frac{(k-1)(k+1)(2k+3)}{6(2k+1)(2k-1)^2} \right]^{1/2}$&
          $-2\left[ \frac{(k-1)(k+1)(2k+3)}{6(2k+1)(2k-1)^2} \right]^{1/2}$ \\
2,$-$2 & 1,0 &$\left[ \frac{(k-1)(k+1)}{2(2k+1)(2k-1)} \right]^{1/2}$ & 0 \\
2,$-$2 & 0,0 &$\left[ \frac{k(k-1)}{3(2k+1)(2k+3)} \right]^{1/2}$ &
          $-\left[ \frac{k(k-1)}{3(2k+1)(2k+3)} \right]^{1/2}$ \\
2,0 & 2,2 &$-\left[ \frac{k(2k-1)(k+2)}{6(2k+1)(2k+3)^2} \right]^{1/2}$ &
           $-2\left[ \frac{k(2k-1)(k+2)}{6(2k+1)(2k+3)^2} \right]^{1/2}$ \\
2,0 & 1,0 & $-\frac{1}{2} \left[ \frac{3}{(2k-1)(2k+3)} \right]^{1/2}$ & 0 \\
2,0 & 0,0 & $-\frac{1}{3} \left[ \frac{2k(k+1)}{(2k-1)(2k+3)} \right]^{1/2}$ &
          $\frac{1}{3} \left[ \frac{2k(k+1)}{(2k-1)(2k+3)} \right]^{1/2}$ \\
2,2 & 1,0 & $-\left[ \frac{k(k+2)}{2(2k+1)(2k+3)} \right]^{1/2}$ & 0 \\
2,2 & 0,0 & $\left[ \frac{(k+1)(k+2)}{3(2k+1)(2k+3)} \right]^{1/2}$ &
           $-\left[ \frac{(k+1)(k+2)}{3(2k+1)(2k+3)} \right]^{1/2}$ \\
\hline \hline
$\alpha =(j_1,j_2)$ & $\beta =(j_1,j_2)$ & 
$C_{\alpha \alpha }^{(-)(2)}$ & $C_{\alpha \alpha }^{(-)(3)}$=
$C_{\alpha \alpha }^{(-)(4)}$ \\
\hline
2,1 & 2,$-1$ & $\frac{1}{2} \left[ \frac{(k-1)(k+2)}{(2k+1)^2} \right]^{1/2}$ &
           $\left[ \frac{(k-1)(k+2)}{(2k+1)^2} \right]^{1/2}$ \\
2,1 & 1,$-$1 & $\frac{1}{2} \left[ \frac{(k+1)(k+2)}{(2k+1)^2} \right]^{1/2}$ &
           0 \\
2,1 & 1,1 & $-\frac{1}{2} \left[ \frac{k(k+2)}{(2k+1)^2} \right]^{1/2}$ & 0\\
2,$-$1 & 1,1& $-\frac{1}{2} \left[ \frac{k(k-1)}{(2k+1)^2} \right]^{1/2}$ & 0\\
2,$-$1&1,$-$1&$\frac{1}{2} \left[ \frac{(k-1)(k+1)}{(2k+1)^2}\right]^{1/2}$&0\\
1,1 & 1,$-1$ & $-\frac{1}{2} \left[ \frac{k(k+1)}{(2k+1)^2} \right]^{1/2}$ & 
           $\left[ \frac{k(k+1)}{(2k+1)^2} \right]^{1/2}$ \\
1 & $-1$ & $ \left[ \frac{k(k+1)}{(2k+1)^2} \right]^{1/2}$ & 0 \\
\hline
\end{tabular}
\caption{}
\end{table}

\begin{table}[t]
\centering
\begin{tabular}{|c|c|c|c|}  \hline
$\alpha =(j_1,j_2)$ & $\beta =(j_1,j_2)$ & $C_{\alpha \beta }^{(+)(5)}$ &
$C_{\alpha \beta }^{(+)(6)}$  \\
\hline  
2,2 & $0$ & 0 & 
       $-\frac{1}{2} \left[ \frac{k(k+2)}{(2k+1)(2k+3)} \right]^{1/2}$\\
2,$-$2 & 0 & 0 & 
      $\frac{1}{2} \left[ \frac{(k-1)(k+1)}{(2k-1)(2k+1)} \right]^{1/2}$\\
2,0 & 0 & 0 & 
      $-\frac{1}{2} \left[ \frac{3}{2(2k-1)(2k+3)} \right]^{1/2}$ \\
1,0 & 0 & $-\frac{1}{\sqrt{2}}$ & $-\frac{1}{2\sqrt{2}}$ \\
\hline \hline
$\alpha =(j_1,j_2)$ & $\beta =(j_1,j_2)$ & $C_{\alpha \beta }^{(-)(5)}$ &
$C_{\alpha \beta }^{(-)(6)}$ \\
\hline  
2,1 & 1 & 0 &
       $\frac{1}{2} \left[ \frac{k(k+2)}{2(2k+1)^2} \right]^{1/2}$ \\
2,1 & $-1$ & 0 &
       $-\frac{1}{2} \left[ \frac{(k+1)(k+2)}{2(2k+1)^2} \right]^{1/2}$ \\
2,$-$1 & 1 & 0 &
      $\frac{1}{2} \left[ \frac{k(k-1)}{2(2k+1)^2} \right]^{1/2}$ \\       
2,$-$1 & $-1$ & 0 &
      $-\frac{1}{2} \left[ \frac{(k+1)(k-1)}{2(2k+1)^2} \right]^{1/2}$ \\         
1,1 & 1 & $\frac{1}{\sqrt{2}}$ &
       $\frac{1}{2} \left[ \frac{k^2}{2(2k+1)^2} \right]^{1/2}$ \\
1,1 & $-1$ & 0 &
       $-\frac{1}{2} \left[ \frac{k(k+1)}{2(2k+1)^2} \right]^{1/2}$ \\ 
1,$-$1 & 1 & 0 &
      $-\frac{1}{2} \left[ \frac{k(k+1)}{2(2k+1)^2} \right]^{1/2}$ \\
1,$-$1 & $-1$ & $\frac{1}{\sqrt{2}}$ & 
      $\frac{1}{2} \left[ \frac{(k+1)^2}{2(2k+1)^2} \right]^{1/2}$ \\
\hline
\end{tabular}
\caption{}
\end{table} 

\begin{table}
\centering
\begin{tabular}{|c|c|c|c|c|}  \hline
n & $\omega^2_n = 1-1/n^2$ & 100 point lattice & 200 point lattice &
400 point lattice\\
\hline
1 & 0      & 0.0008 & 0.0003 & 0.0001 \\
2 & 0.75   & 0.7685 & 0.7687 & 0.7688 \\
3 & 0.8889 & 0.8947 & 0.8948 & 0.8949 \\
4 & 0.9375 & 0.9399 & 0.9400 & 0.9401 \\
5 & 0.9600 & 0.9608 & 0.9614 & 0.9614 \\
6 & 0.9722 & 0.9667 & 0.9685 & 0.9709 \\
7 & 0.9796 & -      & 0.9814 & 0.9782 \\
8 & 0.9844 & -      & -      & 0.9833 \\
9 & 0.9877 & -      & -      & -      \\
\hline
\end{tabular}
\caption{}
\end{table} 


\begin{thebibliography}{99}

\bibitem{Rubakov82} V. A. Rubakov {\it Nucl. Phys.} B203 , 311 (1982);
C. Callan Jr., {\it Phys. Rev. } D25, 2141 (1982); 
C. Callan Jr., {\it Phys. Rev. } D26, 2058 (1982);
B. Sathiapalan and T. Tomaras {\it Nucl. Phys. } B224, 491 (1983);
A. Sen {\it Phys. Rev.} D28, 876 (1983)
\bibitem{Tang82} J. F. Tang {\it Phys. Rev. } D26, 510 (1982)
\bibitem{Tang83} J. F. Tang {\it Phys. Lett. } 120B, 364 (1983)
\bibitem{Marciano83} W. J. Marciano and I. J. Muzinich { \it Phys. Rev. } D28,
973 (1983)
\bibitem{Sonoda84} H. Sonoda {\it Phys. Lett. } 143B, 142 (1984)
\bibitem{Ajithkumar88} C. M. Ajithkumar, M. Raveendranadhan and M. Sabir
{\it J. Phys. G: Nucl. Phys.} 14 ,433 (1988)
\bibitem{Olsen90} H. A. Olsen, P. Osland and T. T. Wu {\it Phys. Rev. } D42,
665 (1990)
\bibitem{tHooft74} 't Hooft {\it Nucl. Phys. } B79, 276 (1974)
\bibitem{Polyakov74} A. M. Polyakov {\it Sov. Phys. } J.E.T.P. {\it Lett. }
20, 194 (1974), [{\it Pisma Zh. Eksp. Teor. Fiz.} 20, 430 (1974)]
\bibitem{Montonen77} C. Montonen and D. Olive {\it Phys. Lett.} 72B, 117 (1977)
\bibitem{Bogomolnyi76} E.B.~Bogomol'nyi, {\it  Sov.~J.~Nucl.~Phys.~} 24, 449
(1976), [{\it Yad.~Phys.~} 24, 861 ]
\bibitem{Seiberg94} N. Seiberg and E. Witten {\it Nucl. Phys.} B426, 19 (1994)
\bibitem{Sen94} A. Sen {\it Phys. Lett.} B329, 217 (1994)
\bibitem{Gibbons86} G. W. Gibbons and N. S. Manton {\it Nucl. Phys.} B274, 183
(1986)
\bibitem{Prasad75} M. K. Prasad and C. M. Sommerfield {\it Phys. Rev. Lett.}
12,760 (1975)
\bibitem{Julia75} B. Julia and A. Zee {\it Phys. Rev.} D11, 2227 (1975)
\bibitem{Carson90} L. Carson, Xu Li, L. McLerran and Rong-Tai Wang,
{\it Phys. Rev.} 42, 2127 (1990)
\bibitem{Goodband95a} M. Goodband and M. Hindmarsh, {\it Phys. Rev. D} 52, 4621
(1995)
\bibitem{Nielsen73} H.B.~Nielsen and P.~Olesen, {\it Nucl.~Phys.~} B61, 
45 (1973)
\bibitem{Rajaraman82}R. Rajaraman, {\it Solitons and Instantons\/} 
{(North Holland, Amsterdam, 1982)}
\bibitem{Yoneya77} T. Yoneya {\it Phys. Rev.} D16, 2567 (1977)
\bibitem{Manton77} N. S. Manton {\it Nucl. Phys.} B126, 525 (1977)
\bibitem{Mandula77} J. E. Mandula {\it Phys. Lett. } 67B, 175 (1977)
\bibitem{Olive95} D. Olive {\tt hep-th/9508089}
\bibitem{Manton85} N. S. Manton {\it Phys. Lett.} 154B, 397 (1985)

\end{thebibliography}
\end{document}